\documentclass{nime-alternate} %
\usepackage{anonymize}        %
\usepackage[utf8]{inputenc}

\begin{document}

\conferenceinfo{NIME'20,}{July 21-25, 2020, Royal Birmingham Conservatoire, ~~~~~~~~~~~~ Birmingham City University, Birmingham, United Kingdom.}
\title{A Laptop Ensemble Performance System using Recurrent Neural Networks}

\label{key}
\numberofauthors{2}
\author{
\alignauthor
Rohan Proctor\\
       \affaddr{Research School of Computer Science}\\
       \affaddr{Australian National University}\\
       \affaddr{Canberra, Australia}\\
       \email{rohan.proctor@anu.edu.au}
\alignauthor
Charles Patrick Martin\\
       \affaddr{Research School of Computer Science}\\
       \affaddr{Australian National University}\\
       \affaddr{Canberra, Australia}\\
       \email{charles.martin@anu.edu.au}
}

\maketitle

\begin{abstract}
The popularity of applying machine learning techniques in musical domains has created an inherent availability of freely accessible pre-trained neural network (NN) models ready for use in creative applications. This work outlines the implementation of one such application in the form of an assistance tool designed for live improvisational performances by laptop ensembles.
The primary intention was to leverage off-the-shelf pre-trained NN models as a basis for assisting individual performers either as musical novices looking to engage with more experienced performers or as a tool to expand musical possibilities through new forms of creative expression. The system expands upon a variety of ideas found in different research areas including new interfaces for musical expression, generative music and group performance to produce a networked performance solution served via a web-browser interface. The final implementation of the system offers performers a mixture of high and low-level controls to influence the shape of sequences of notes output by locally run NN models in real time, also allowing performers to define their level of engagement with the assisting generative models.
Two test performances were played, with the system shown to feasibly support four performers over a four minute piece while producing musically cohesive and engaging music. Iterations on the design of the system exposed technical constraints on the use of a JavaScript environment for generative models in a live music context, largely derived from inescapable processing overheads. 

\end{abstract}



\section{Introduction}

A side effect of the wealth of research being undertaken in intelligent musical systems has been the advent of freely available pre-trained machine learning models for music. This work presents the design and implementation of a proof-of-concept assistance tool for use by laptop ensembles in live improvisational performances. The system utilises off-the-shelf deep learning models offered by the Magenta \cite{Roberts:2019aa} project to assist performers in creating musically consistent compositions with respect to the ensemble during performances. We discuss the experience of overcoming the technical challenges and performance constraints of adopting off-the-shelf machine learning music generation into a browser-based laptop ensemble environment. While synchronisation of multiple ML-generated music models in real-time threatened the feasibility of this project, we developed successful workarounds that were successful in enabling two concert performances. This work contributes a practical design for an ensemble performance system using recurrent neural networks. We have shown this system to be appropriate for applying ML music generation in collaborative performance and used this system to explore the musical practices afforded by current work in musical ML.

\begin{figure}
  \centering
  \includegraphics[width=\columnwidth]{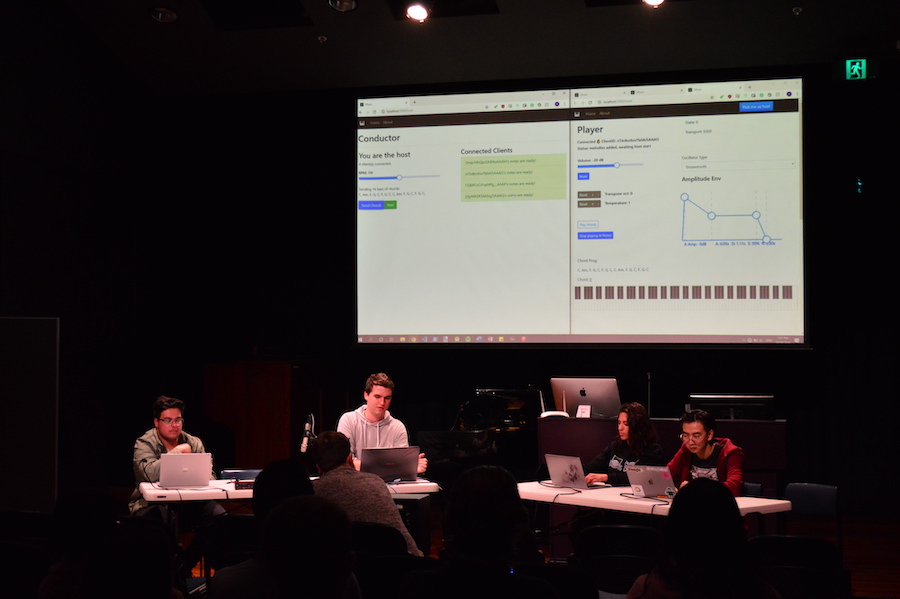}
  \caption{Our ensemble RNN performance system in concert. This system
  runs in web browsers and allows a ``conductor'' to control the tempo
and chord progression of a song, while a number of ``performers''
steer independent RNN models that generate melodies from the chords.
The system works in real-time for collaborative performances.}
  \label{fig:performances}
\end{figure}

The focus of the design of the system was to utilise note sequence generation functionality of musical neural network models to assist performers in selecting musically coherent notes to play, allowing less musically experienced performers to fearlessly contribute to a laptop ensemble during a performance alongside more experienced players. As the pre-trained model checkpoints offered by the Magenta project are available for use through a JavaScript API, the system was developed as an application targeting web browsers, also providing added benefits of inherent accessibility and minimal set-up time. Performers running individual neural network models locally can choose to fully incorporate, borrow from or ignore generative output offered by the models in expressing musical ideas during performances. This interaction is designed to align with performers’ respective levels of musical knowledge or experience and can additionally be used as a creative tool for musical expression. The system was used in two performances, ultimately proving to be successful in delivering a final four minute improvised piece, supporting four performers to create an aesthetically interesting and musically consistent performance.

\section{Background}

\subsection{Generative Music in Group Performance}

Approaches using generative techniques within musical ensembles from existing literature can be generally classified based on the type of interaction intended between generative models and human performers. On one end of the spectrum, approaches like the Logos Robot Orchestra \cite{logos} prevent human performers from influencing the output of any generative model with the basis of the performance interaction becoming a human accompaniment of the automated instruments or vice versa. On the other side, strategies aim to blend the human-model interaction by either allowing generative models to borrow, learn from or listen to human output to shape their own output in response \cite{chen_2017} or providing humans the capacity to explicitly influence musical form of generative agents \cite{Kapur:2011it}.

Many implementations of generative music systems in ensembles offer different types of high control for humans during performances. LolBot \cite{Subramanian2012LOLbotMM} is an autonomous program which collaborates with human performers in a laptop ensemble environment. In a performance, it functions by analysing a library of human-made rhythmic patterns while ‘listening’ to the current patterns being played to generate new, fitting patterns to play. A notable feature of this tool is that it enables users to interact with the generative model with high level controls, to define the level of rhythmic coherence or contrast that newly generated patterns should have relative to patterns being played \cite{Subramanian2012LOLbotMM}.
Thom’s Band Out of Box \cite{thom} is structured around an agent responding to human improvisation in a call and response manner. In performing alongside a human, the agent aims to strike a balance between operating within musical constraints established by the human’s performance and contributing novel musical ideas that remain cohesive within these musical bounds. Other works such as the Inmamusys system \cite{Imma} push the creative extent of the method of human control over generative models, in this instance using collaborative agents to synthesise human emotions into musical compositions.

\subsection{Generative Models and Musical ML}

Generative models for the creation of music have developed hand in hand with technological change over time. Hiller’s 1959 Illac Suite represents the first computational approach, using Monte Carlo sampling and Markov Chain state transitions to the create musical scores \cite{Hiller:1979:EMC:578548}.
Brian Eno’s 1978 album Ambient 1: Music for Airports is a notable example of the use of heuristic models used for musical expression, using loops of magnetic tape of varying lengths played simultaneously to produced different sequences of notes at tape-length-dependent intervals forming rarely repeated cycles of notes \cite{eno1996year}.

The advent of machine learning has impacted not only generative composition of music but also audio synthesis \cite{GANSynth}, audio processing \cite{Pardo:2012:BPA:2390848.2390852} and sound design \cite{Miranda0}. Recurrent Neural Networks (RNNs) with Long Short-Term Memory (LSTM) units have become widely used for generating different kinds of music from improvising the blues \cite{Eck} to creating aesthetically plausible pop music \cite{chu2016song}. These neural network models have also been shown to be applicable to real-time musical interaction~\cite{Martin2019, Naess:2019aa}.

The Magenta project is a Google-led effort to widen the availability of machine learning based models for use in new practical applications \cite{Roberts:2019aa}. The project represents a significant step in delivering tangible hooks into machine learning research to the masses by offering an open source implementation of different Recurrent Neural Network (RNN) and Variational Auto Encoder (VAE) architectures along with a freely available set of pre-trained model checkpoints for use. Access is provided through the Majenta.js API built upon the Tenserflow.js API.

\subsection{Collaborative Music Systems}

The diversity of approaches in collaborative computer music has led to a variety of strategies to overcome common problems. 
Timing control within implementations of generative musical agents in ensembles can be problematic, often due to either a lack of perception of temporal scope, computational processing time overheads, signal delay over networks or an inability to adjust or match tempo in real-time. Solutions to ensuring synchronisation found in literature have included real-time adaptive tempo systems \cite{Robertson:2013:SSS:2492334.2492340}, forward-looking predictive rhythmic approximations over networks \cite{Sarkar:2007aa}, performance delays as-a-feature \cite{ninjam} or the minimisation of network latency by abstracting performer output to less data-intensive formats including MIDI \cite{lazzaro}.

Implementations of complex musical systems commonly simplify interaction for performers such that they can intuitively begin using a novel system. For example, \cite{Naess:2019aa} presents a system where a person can control the musical form of a melody by turning two physical knobs on a self-contained instrument. The knobs adjust the temperature of an underlying RNN as well as the audio volume of the notes played. Abstracted levels of control can be leveraged for new creative expressiveness, such as the use of hand gestures \cite{beyer} or the manipulation of tethered strings \cite{wang_bryan_oh_hamilton_2009} to produce audio.
In ensemble environments, control schemes can range from one-to-one control over analogous ‘instruments’ \cite{Kapur:2011it} to the sharing of live ‘coded’ symbolic text used to express musical intent \cite{Subramanian2012LOLbotMM}. Restriction of performer control within ensembles \cite{martin:2017} can also be used as a method of ensuring musical consistency or direction during performances.

\section{Design and Implementation}

This system was designed to explore the capacity for ensemble music-making using off-the-shelf RNN models. The intended performance group was the ANU Laptop Ensemble, a collective of undergraduate and master's students studying computer science and music. The system was created with the following design goals in mind:
\begin{description}
\item[Live Generation and Collaboration.] The system needed to support three or more performers in a dependable and stable manner with music generation responding in real-time to the performers commands.
\item[Maintenance of creative control.] The tool should allow the performers to choose a level of control over the model output.
\item[Musicality.] The system should sound appropriately musical and maintain synchronised rhythm.
\item[Accessibility.] The system should be straightforward to run and interact with for all members of the ensemble.
\end{description}
The system comprises a central server with which a single \emph{conductor} client and multiple \emph{performer} clients communicate via a WebSocket API during a performance. One member of an ensemble can run the server and host client (as well as a performer client) simultaneously over WLAN from which all other performers connect to via a single IP address in a browser window.
The entire system is implemented in JavaScript, targeting the use of the Magenta.js API, using Tone.js for audio synthesis and sequencing and NodeJS to run the server.

\begin{figure}
  \centering
  \includegraphics[width=\columnwidth]{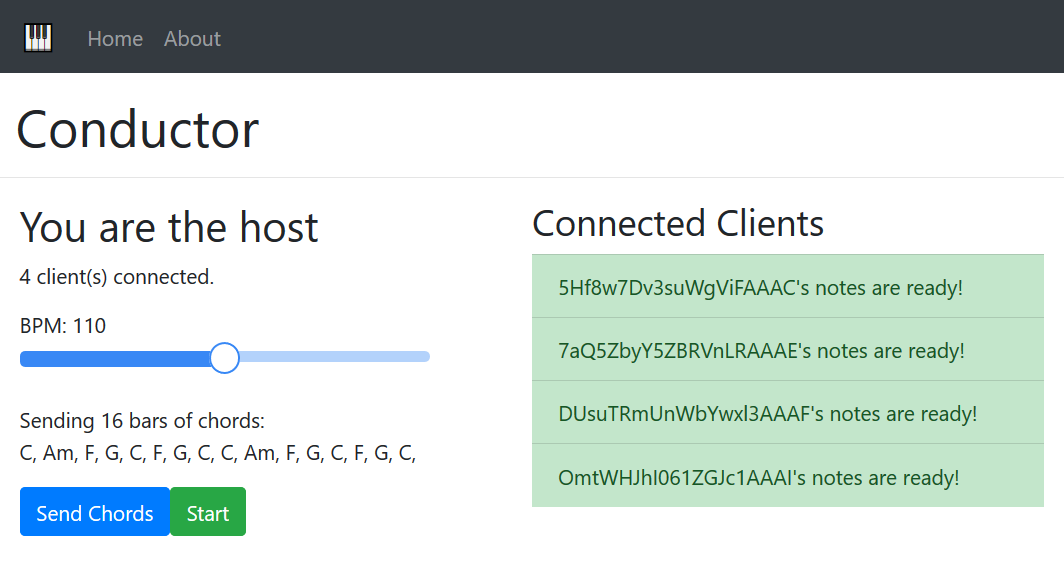}\\
  \includegraphics[width=\columnwidth]{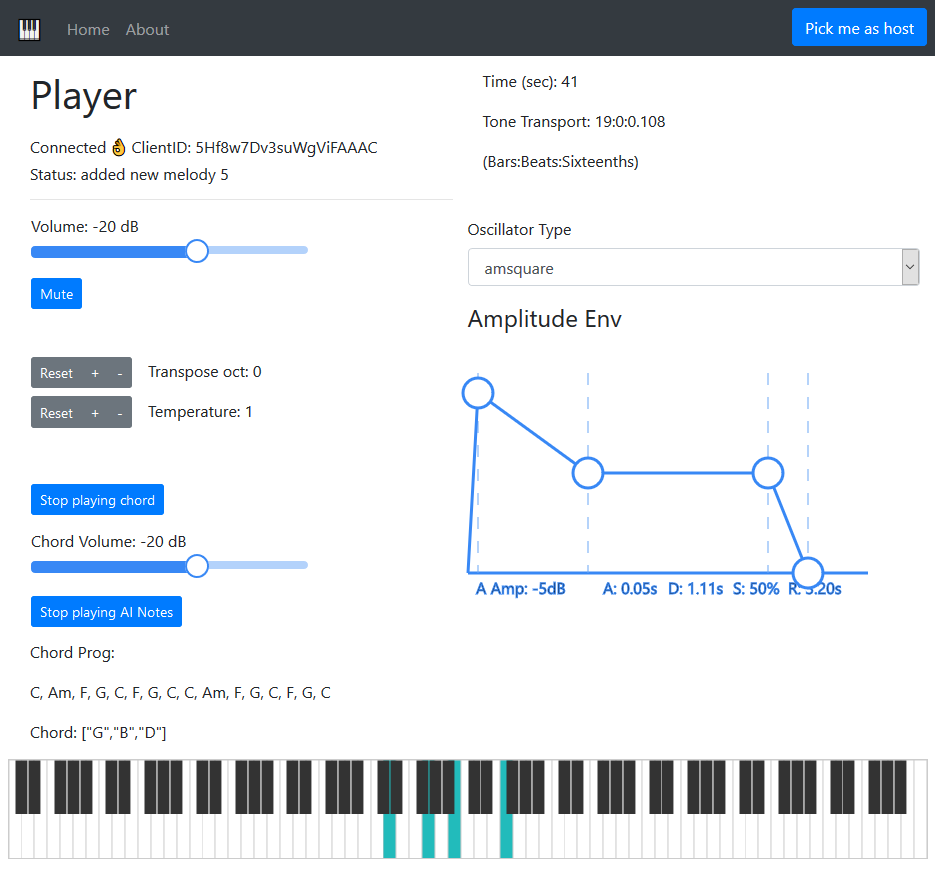}
  \caption{The conductor (top) and performer (bottom) client user interfaces.}
  \label{fig:host-performer-clients}
\end{figure}

\begin{figure}
  \centering
  \includegraphics[width=\columnwidth]{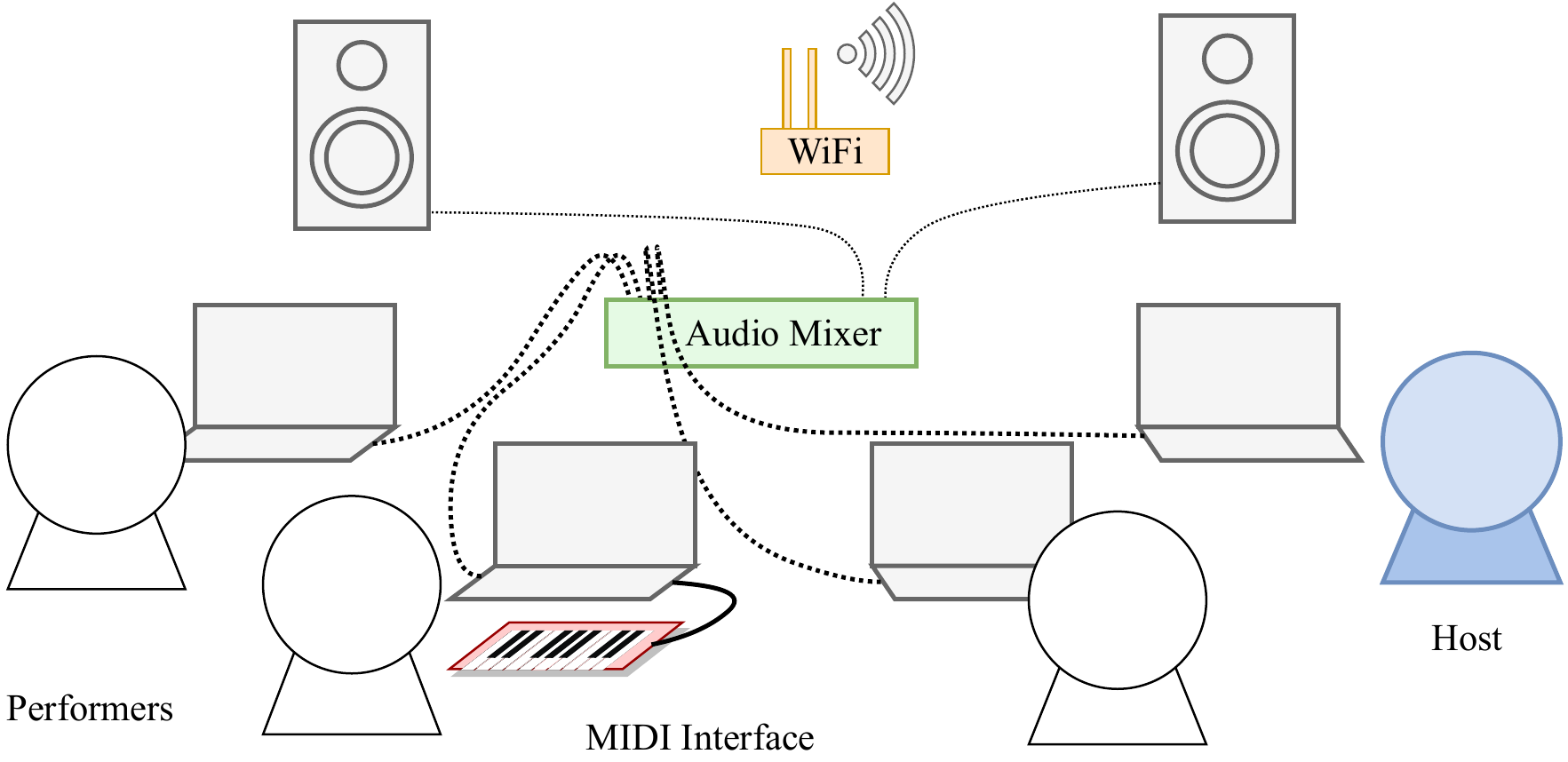}
  \caption{Performance set-up}
  \label{fig:performance-diagram}
\end{figure}

\begin{figure}
  \centering
  \includegraphics[width=\columnwidth]{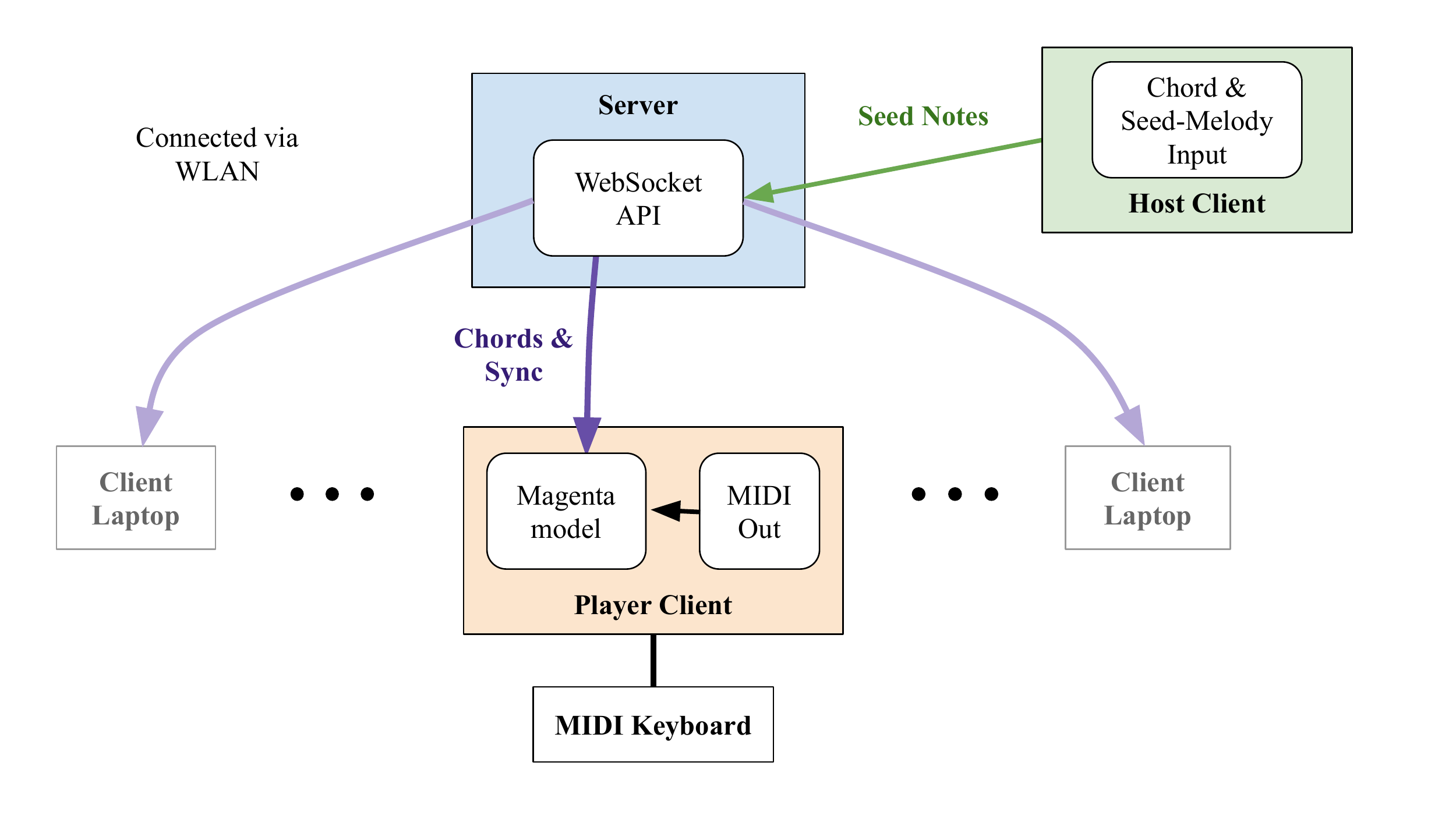}
  \caption{System overview with MIDI interface}
  \label{fig:system-diagram}
\end{figure}

\subsection{Performance System}

The host client acts as a central conductor for all performer clients. During an initialisation handshake procedure with each performer client, the host communicates a seed chord progression, melody and tempo before the commencement of a performance, shown in Figure \ref{fig:system-diagram}.
Each performer client runs a local RNN model using the Magenta API which receives the initial chord progression and seed melody from the host and generates two initial 16 bar sequences of notes using a pre-selected, pre-trained model checkpoint available from the Magenta project website. The model checkpoint used in both performances was \textit{chord-pitches-improv}  \footnote{Available: \url{https://goo.gl/magenta/js-checkpoints}}, chosen due to the capacity to input both a melody and underlying chord progression to ensure that generated notes adhere to a level of musical structure, maintaining musical cohesiveness between performers.

Upon commencement of a performance by the host performer, the server begins a central timer and communicates a start message to all clients which initiates audio output. Note sequences are continuously generated by each performer client, using the previous 16 bars of notes played by the performer as input to local RNN model for the generation of a new 16 bar sequence of notes. One significant caveat of this approach is that the generation a sequence of notes comes at a significant processing time cost as outlined in section~\ref{sec:StaggeredGen}.

\subsubsection{Staggered Note Sequence Generation}

\label{sec:StaggeredGen}
The processing time associated with generating a new sequence of notes for a performer client presented a significant challenge in maintaining flow in a live music performance context. The processing time results in a complete freeze of the performer client audio output and interface for a few seconds, dependant on the hardware in a performer's laptop. Unfortunately, the freeze time proved to be inescapable for performer clients, with no concurrency-based escape available due to Javascript's WebWorkers not allowing access to the AudioContext in the base thread as required by the MagentaJS API. 

An alternative solution to this problem \cite{chen_2017}, is to send recorded midi data to an external server which performs the task of generating a sequence then returns note sequence to be played in a seemingly seamless manner. This is a solid solution for the use case of only running one model but becomes unfeasible in the context of a live group performance where a server instance would have to be run for each performer, especially in an environment where a significant amount of network traffic is present. This would also increase the amount of external dependencies and introduce the need for an internet connection during performances.

A potential solution may lie in experimental additions of AudioWorklets to Google Chrome\footnote{\url{https://developers.google.com/web/updates/2017/12/audio-worklet}} and Mozilla Firefox\footnote{\url{https://developer.mozilla.org/en-US/docs/Web/API/AudioWorklet}} which could potentially assist in alleviating this issue, however have not been tested as part of this paper.

\begin{figure}
    \centering
    \includegraphics[width=\columnwidth]{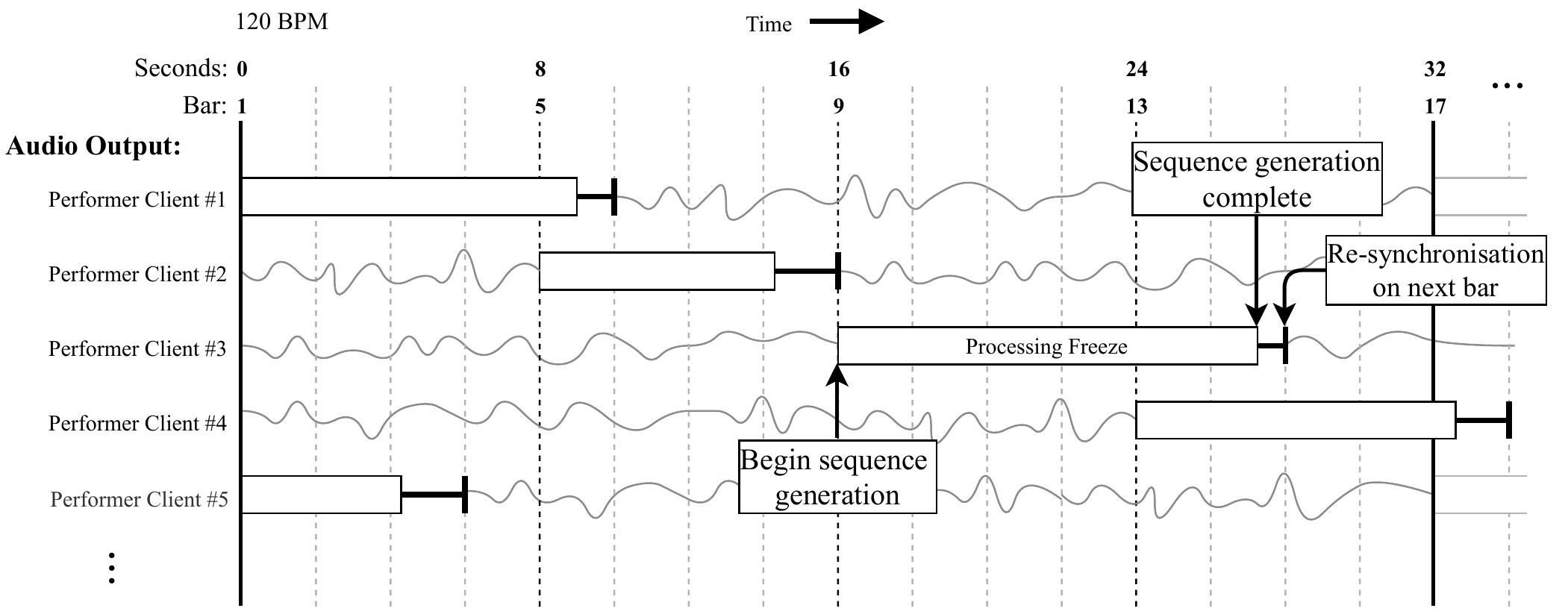}
    \caption{Timeline of sequence generation}
    \label{fig:sequence_generation}
\end{figure}

As a result, a system for keeping all performer clients on track was developed, ensuring continuous audio output by dynamically staggering the timing of the generation of note sequences for each performer client across an ensemble. As shown in Figure \ref{fig:sequence_generation}, the system accounts for the time spent frozen by each performer client and re-synchronises them to all other clients using the server clock to insert them back into the performance at the correct moment.

\subsection{Performer Interaction}

Each performer is provided with an interface akin to a simplified DAW shown in Figure \ref{fig:host-performer-clients}, providing controls for synthesizers driven by ToneJS producing audio as well as high level controls over the local RNN model including model temperature and transposition of output melodies. The interface additionally provides timing information and shows the current underlying chord and overall chord progression such that a musically experienced performer can easily play along without using any model-generated notes. Figure \ref{fig:performance-diagram} shows a typical ensemble configuration for performance using optional external MIDI interfaces. 

The implementation permits performers choice over their level of interaction with the assisting model. Performers can choose to fully improvise their output, preventing their local model from playing any notes. This represents the ‘default’ state for any performer in a traditional improvisational ensemble, for example, a trombonist improvising in a jazz quartet. Conversely, they can also take a high-level, hands-off approach and let the model produce all the notes to play while retaining control of the synthesis process to shape the sound produced.
Perhaps the most interesting and creative use of the system lies between these two extremities. Performers can use the temperature control and transpose functions to shape note sequences generated by a model and utilise this as a creative tool. As an example, a performer may reduce the temperature of their model to produce a compositionally simple, mildly-repetitive note sequence within which the notes mostly reflect the root notes of the underlying chord progression with some variation. By transposing this output downwards by a few octaves, selecting a fitting oscillator waveform and appropriate amplitude envelope, a bassline can be generated which can be improvised on top of. In a similar vein for another technique, the temperature can be dramatically increased with the synthesizer volume reduced, the note sequence transposed upwards and the envelope set with small attack and long release to create a randomly 'arpeggiated' wind-chime like effect.
This represents an interesting emergent behaviour of the system which can be leveraged for novel musical expression. Coupled with the control provided to a performer over the digital synthesis process, the system can produce a diverse range of different compositions and sounds.

\section{Performances and Evaluation}

This system was deployed in two live performances with an iteration on the system design after after the first. We describe the performances and evaluate the experience in terms of our design criteria.

\begin{figure*}
  \centering
  \includegraphics[width=0.89\columnwidth]{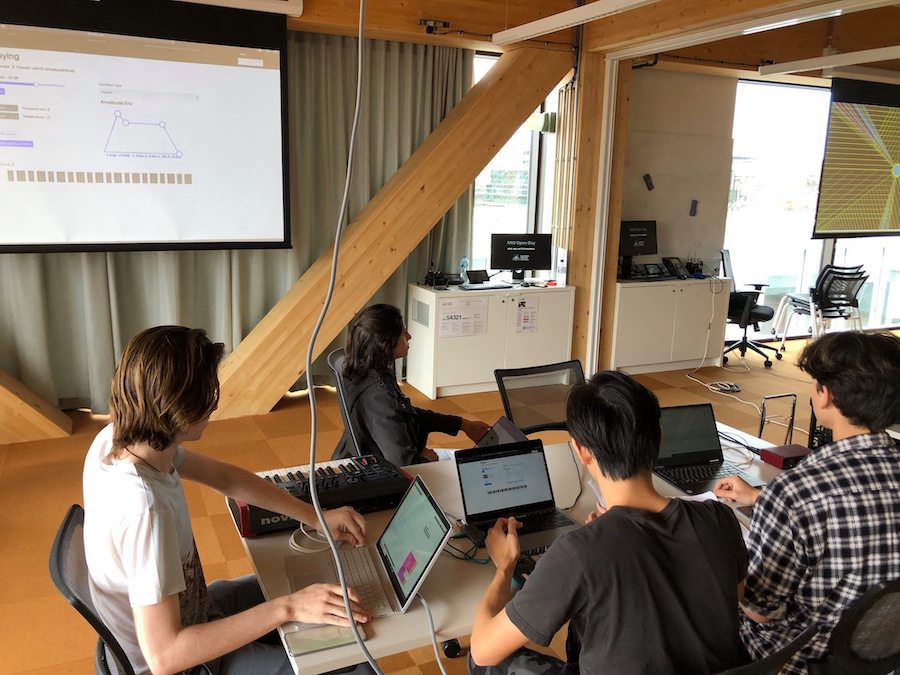}
  \includegraphics[width=\columnwidth]{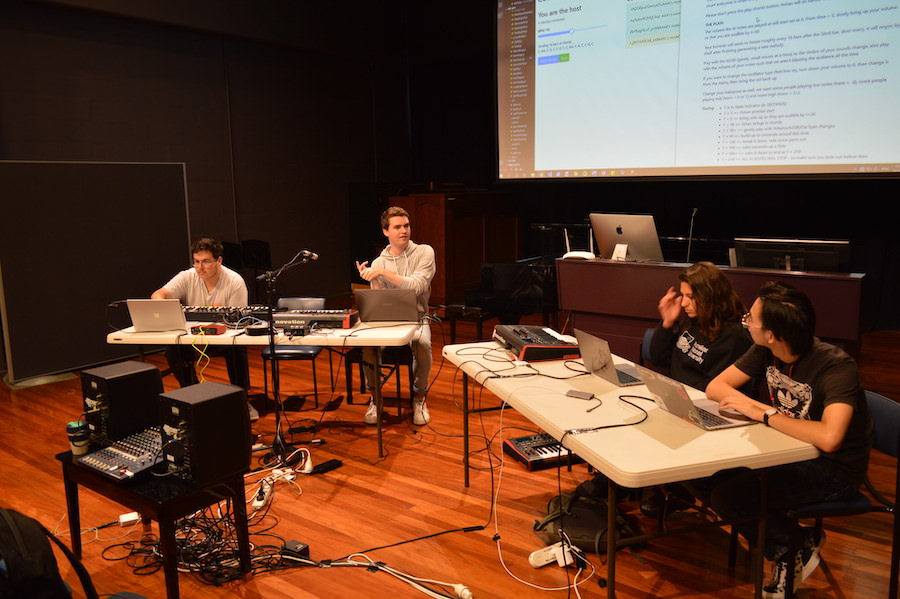}
  \caption{Performances with the laptop ensemble performance system:
    Performance 1 (left) was held during a university open day in a
    classroom; Performance 2 (right) was during a concert of the
    ANU Laptop Ensemble.}
  \label{fig:performances}
\end{figure*}

\subsection{Performance 1}

The first performance using this system was by four members of a laptop ensemble during a university open day (the performance is shown in the left side of Figure \ref{fig:performances}). The performance took place mid-way through the development of the assistance tool and as such, the prototype had a bare minimum feature set, enough for a brief, musically simplistic performance. 
Unfortunately, the performance suffered a breakdown resulting from connection drop outs to the server over the WiFi network and thus was cut short to around two minutes. The breakdown may have been due to using an older WiFi router in a busy university campus environment, and resulted in failures of connections and message sending.
After multiple attempts of refreshing the system, the host client eventually successfully received confirmation from all player clients that their respective NN models had loaded and were ready to begin playing. After three note sequence generation cycles, roughly sixty-four bars into the piece, the connections once again dropped out. The implementation of the performer client at that stage allowed for the continuation of the model loading cycles without any dependency on the server clock. While this behaviour allowed the performance to continue even without network communication, the loss of synchronisation and confusion between performers based on the unexpected behaviour forced a quick end to the piece.

\subsection{Performance 2}

The second performance (see the right side of Figure \ref{fig:performances} and the video available online\footnote{Performance video:\\ \url{https://youtu.be/0nzNdM5RCjY}}) consisted of four (different) performers playing a four minute piece as part of a laptop ensemble concert. One of the key differences between this performance and the first was the use of a plan to define an overall musical structure to the piece. The plan provided a high-level overview of the how the 'energy’ of the music should change over the piece. This was written in text at the bottom of the performer client interface as a list of instructions, for example: ‘@Bar 120: begin to lower the intensity until soft by bar 150’.

Using a more mature version of the tool, the performance remained stable throughout although some unexpected behaviour in non-mission-critical components did occur. One instance of this was an error in rendering the server clock on the performer clients, where the clock displayed a ‘0’ instead of the correct time. This did not affect the model loading cycle procedure, meaning that it would have occurred due to an issue with the client-side React UI rendering rather than the server clock itself. One performer experienced a note getting stuck in the ‘on’ position towards the end of the performance, likely caused by a failure to trigger the note release condition due a sequence generation processing freeze. 

\subsection{Discussion and Lessons Learned}

The two performances gave us the opportunity to improve the stability of the performance system and to understand how it performs in terms of our design criteria of feasibility, creative control, musicality, and accessibility. 

The final implementation of the assistance tool was proven to be feasible for use in live performances with at least four performers. All music was generated by the system in apparent real-time with a range of synthesis controls exposed for performers to contribute expression within performances. Performers could produce musical ideas and respond to one another just as can be expected from a traditional ensemble environment. A few caveats remain with regards to the quality of the interaction. The processing freeze resulted in a loss of control for the performers despite our best efforts to streamline sequence generation---a significant hurdle in presenting a smooth experience. Given the complex nature of the project, working in a limited time-frame as well as some technical constraints and the central target of delivering live performances, the acceptance of this method of operation was decidedly unavoidable.

The system allowed performers to maintain creative control, although balance between performer and RNN control existed in three modes rather than a continuum. Performers could choose to either avoid the use of the models altogether, incorporate them into their creative output while still playing notes themselves or fully rely on the generated melodies while retaining control of the musical voicing and sound design elements. These categories of engagement did support the intention for novice musicians to be able to expressively contribute alongside more experienced performers in a meaningful way. 

Although these different modes of engagement do exist, the question remains if they were all feasible to use in a performance context. For the second performance, all performers showed a heavy reliance on the generated note sequences for the composition of their output, opting to control and shape the sound with the higher level controls available. The models collectively produced a saturation of notes being played simultaneously across the ensemble such that performers had to adjust by controlling their volume levels to thin the musical texture.

In terms of musicality, the note sequences generated by the Magenta models were limited in terms of complexity. 
Potentially as a result of this or alternatively through intended design, the musical output of individuals across the ensemble did remain structurally cohesive and musically ‘appropriate’ \cite{thom} throughout both performances. Melodies produced by the models were not dissonant across the ensemble and did contain a level of musical novelty and interest. As generated note sequences were quantised (by the Magenta model) at intervals with very little variation within a sequence, output melodies tended to lack a human feel due to the inhuman perfection of timing. Chord voicings were very basic, with almost no variation on the playing of continual major triads.

The musical result of this, when using major chord progressions as in both performances, was stylistically akin to a jolly nursery-rhyme. Individual melodies contained recurring musical motifs with some variation dependent on the sampling temperature selected by performers. 
Breaks in audio from the model note sequence generation freeze were audible and apparent in performances. Synchronisation was generally well maintained across all performers, with the exception of some small but noticeable deviations in the first half of the second performance. A variety of simple electronic timbres could be produced through the selection of waveforms available to use with the software based oscillators when coupled with the manipulation of oscillators’ amplitude envelopes.

The system was accessible for the performers in terms of operation during a performance. This was evidenced by the fact that the second performance was delivered after only two runs of the piece by a performing group with no previous experience using the tool (with the exception of the first author). From a designer’s perspective it was only necessary to convey a minimal amount of instruction for a performer to feel comfortable working in a networked environment with a local RNN model. Anecdotally, the largest difficulty was navigating the connection procedure to the central server before the performance began. The system was highly accessible in terms of being a plug and play browser-based solution, requiring no installation or set-up time for performers and minimal effort for the ’host’ performer. 

\section{Conclusions}

This research introduces a novel implementation of a collaborative live performance system utilising recurrent neural network models to musically assist performers and establish new avenues of creativity through a new interface for musical expression. This design applies popular existing RNN models into a web-based laptop ensemble performance system. Laptop performances present particular challenges in terms of rhythmic and harmonic synchronisation, and the human challenges of running software on multiple computer systems. This work has advanced existing deep neural network-based generative music systems, which have, in general, been focussed on composition and individual performance, by encapsulating the generative models in a practical system and addressing some of the challenges of collaborative generative performance.

The system was demonstrated in two concerts, proving to be successful in offering stable, feasible, creatively engaging and musically coherent performances. 
Through a networked performance model, performers were able to effectively collaborate and define their level of engagement with assisting agents through three different modes: rejecting it, relying on it or blending between both means. The use of human-controllable agents for the live generation of reactive note sequences opened new avenues for creative expression and musical performance. Our system could be extended into more specific compositions for generative ensemble, or with more advanced generative music models, as these become available.

While we found that our system was practical during our laptop ensemble performance. Our work exposed technical issues that are critical to the success of ensemble performances in the browser, and applying neural network predictions in real-time. The use of JavaScript and targeting web-browser environments enforced strict limitations in terms of performance overhead, concurrency and overall user experience. 
Given that the sounds cannot be simultaneously generated by the RNN and synthesised in JavaScript, our note freeze work-around was necessary to make use of the Magenta.js models in real-time performance. The resulting pauses and blocking in the user interface was unfortunately noticeable and annoying to the performers. In future, this could potentially be addressed using either externalised server-side processing or currently experimental audio-functional JavaScript Web Workers. Other limitations included the time required to initialise the model and share seed chords and melody, and that seed melodies and chords are presently hard-coded in the host client.

There is much scope for further developments of this work, for instance, by adding an audio effects chain, increasing network stability, or developing a mobile-targeted client. These additions would open up new performance possibilities. The test performances suggested some immediate improvements, for example, an automatic volume fader which anticipates an incoming sequence generation freeze and reduces the volume, increasing back to normal it once the freeze is complete. This would assist greatly in reducing the sometimes jarring nature of the reintroduction of a performer to the performance.

\subsubsection*{Acknowledgments}
We wish to thank the ANU Laptop Ensemble for participating
  in live performances with our system.

\bibliographystyle{abbrv}
\bibliography{references} 
\end{document}